\documentclass[structabstract]{aa}
\usepackage{graphicx}
\usepackage{txfonts}
\usepackage{longtable}
\usepackage{natbib}
\bibpunct{(}{)}{;}{a}{}{,}

\begin{document}

\title{The K supergiant runaway star HD~137071\thanks{Based on observations collected at the Centro Astron´\'mico Hispano Alem\'an (CAHA) at Calar Alto, operated jointly by the Junta de Andalucía and the Instituto de Astrofísica de Andalucía (CSIC)}
}
\author{F. Comer\'on\inst{1}
\and F. Figueras\inst{2}
}
 \institute{
  European Southern Observatory, Karl-Schwarzschild-Str. 2, D-85748 Garching bei M\"unchen, Germany\\
  \email{fcomeron@eso.org}
  \and
  Institut de Ci\`encies del Cosmos (IEEC-UB), Barcelona, E-08028, Spain
  }
%
%
\date{Received; accepted}
\abstract
{Extensive work exists on runaway massive stars having peculiar motions much higher than those typical of the extreme Population~I to which they belong. Work on runaways has focused almost exclusively on O and B stars, most of which undergo a red supergiant phase before ending their lives as supernovae. Very few examples are known of red supergiant runaways, all of them descending from the more massive O-type precursors, but none from the lower mass B-type precursors, although runaway statistics among B-type stars suggest that K-type runaways must be relatively numerous.
}
{We study HD~137071, a star that has been considered so far as a normal K-type red giant. Its parallax measured by Gaia and the derived luminosity suggest that it is actually a supergiant, whereas its derived distance to the galactic plane and its spatial velocity of $54.1$~km~s$^{-1}$ with respect to the local standard of rest suggest that it is also a runaway star. However, intrinsic limitations in determining the trigonometric parallaxes of cool supergiants, even in the Gaia era, require accurate spectral classifications for confirmation.
}
{We present visible spectroscopy obtained with the 2.2m telescope at Calar Alto Observatory and compare it with the spectra of MK standard stars to produce an accurate spectral classification, including the determination of its luminosity class. We complement this information with astrometric data from the Gaia DR2 catalog.}
{We reliably classify HD~137071 as a K4II star establishing its membership to the extreme Population~I, which is in agreement with the luminosity derived using the Gaia DR2 parallax measurement. Kinematical data from the Gaia DR2 catalog confirm its high spatial velocity and its runaway nature. Combining the spectral classification with astrometric information, a state-of-the-art galactic potential model, and evolutionary models for high-mass stars we trace the motion of HD~137071 back to the proximities of the galactic plane and speculate on which of the two proposed mechanisms for the production of runaway stars may be responsible for the high velocity of HD~137071. The available data favor the formation of HD~137071 in a massive binary system where the more massive companion underwent a supernova explosion about 32~Myr ago.}
{}



\maketitle

\section{Introduction} \label{intro}

The large fraction of runaways among the most massive stars continues to pose challenges, both to obtain reliable, unbiased statistics \citep{Stone91,Maiz18} and to explain their origin in terms of the supernova disruption channel \citep{Blaauw64} or the dynamical ejection channel \citep{Poveda67} that were initially proposed over half a century ago to account for the their large spatial velocities \citep{Portegies00, Fujii11, Perets12}. The potential of the Gaia astrometric mission to improve the statistics of runaway stars has been demonstrated by several recent studies \citep{Maiz18, Fernandes19, Rate20}, but much work remains to be done to obtain complete statistics over the whole massive star range that should form the basis for comparison with the predictions of simulations of the mechanisms accounting for their origin.

Nearly all the observational studies of massive runaway stars have focused on stars of the O and early B types. With the exception of the most massive ones, all those stars undergo the red supergiant phase before ending their lives as supernovae, and therefore we should expect runaway red supergiants to be also relatively abundant. However, very few examples are known. A literature search reveals only Betelgeuse \citep{Harper08}, $\mu$~Cep \citep{Cox12} and IRC~10414 \citep{Gvaramadze14} as the three known runaway supergiants in our galactic neighbourhood. An extragalactic member located in M31 has been recently added to the sample \citep{Evans15}. All these are luminous M-type supergiants with very massive, short-lived O-type progenitors, and very little is known about runaways among less luminous K supergiants, which are the descendants of early-type B stars. This is partly due to the ease with which red supergiants can be misclassified as red giant branch stars, which are much more abundant and very similar both photometrically and spectroscopically. They are also much older, therefore having velocity dispersions comparable to the velocities of actual runaway stars. In principle, accurate distance determinations should help in separating giants from supergiants on the basis of their luminosities. \citet{Messineo19} have studied a large sample of cool supergiants characterized using data from the Gaia DR2 catalog \citep{GaiaDR218}. However, cool supergiants are intrinsically difficult objects for astrometry because their huge convective cells can move the photocenter randomly over the disk of the star with amplitudes exceeding their parallax, as dramatically demonstrated by interferometric images of Betelgeuse \citep{Lopez18}, thus intrinsically limiting the accuracy with which the parallax can be determined regardless of their distance.

Red supergiant runaways, being older than their main sequence precursors, make it possible to investigate old ejection episodes. Furthermore, evolutionary models constrain the ages at which massive stars enter the red supergiant phase, and the relatively short duration of that phase leads to a fairly tight mass-luminosity-age relationship. Its scatter is mainly due to the rotational velocity of the precursor \citep{Maeder00, Ekstroem12} and, in close binary systems, by earlier mass transfer episodes between the star and a massive companion \citep{vanBeveren98}.

 In this paper we discuss evidence for a nearby runaway supergiant, \object{HD~137071}, which may be the second nearest runaway supergiant after Betelgeuse, the only K supergiant identified thus far, and therefore one of the oldest massive runaway stars known. We present spectroscopy that confirms HD~137071 as a luminosity class II star\footnote{Traditionally, stars of luminosity class II are referred to as {\it bright giants} rather than {\it supergiants}, a term that is reserved for luminosity class I alone in the phenomenological MK classification scheme. We adopt here the evolutionary, more physically motivated definition of red supergiant as the stage of a massive star on its way to end its life as a core-collapse supernova in which it is burning helium at its core under non-degenerate conditions. With this definition and the derived characteristics of HD~137071, the separation between supergiants and bright giants becomes rather arbitrary, and we therefore refer to HD~137071 as a supergiant to stress its physical commonalities with its more massive counterparts spectroscopically classified as belonging to luminosity class I.}, and we use Gaia DR2 astrometry and evolutionary models to carry out a tentative reconstruction of its history allowing us to speculate on its mechanism of origin and its consequences.

\section{HD137071 \label{hd137071}}

HD~137071 is a relatively bright ($V = 5.5$), red star in the constellation Bootes, far removed from the galactic equator at a galactic latitude $b = +56^\circ 363$. No dedicated studies exist on HD~137071, but it has appeared in the literature as a part of large samples of giants, with a K4III or K5III classification. Subsolar metallicities have been reported by \citet{McWilliam90} ([Fe/H]$\sim -0.31 \pm 0.15$) and \citet{Cenarro07} ([Fe/H]$\sim -0.21 \pm 0.1$), but more recent work of \citet{Prugniel11} using the state-of-the-art MILES code yields a slightly supersolar metallicity [Fe/H]$= +0.09 \pm 0.05$.

\citet{Tetzlaff11} first identified HD~137071 as a candidate young runaway using Hipparcos distances and proper motions for a large sample of stars whose youth was suspected from the fit to their positions in color-absolute magnitude diagrams, in a study not restricted to any particular range of spectral types. In parallel, we have identified HD~137071 as a supergiant candidate based on its 2MASS $J$ and $K_S$ magnitudes and its Gaia~DR2 parallax in the course of an ongoing study of cool supergiants in the solar neighbourhood. Its high galactic latitude, implying a vertical distance of $524^{+43}_{-38}$~pc over the galactic plane, and its peculiar velocity of $54.1$~km~s$^{-1}$ are both unexpected for a normal cool supergiant belonging to the extreme Population I, and even for a massive RGB star having a relatively short-lived progenitor with an initial mass of a few solar masses, which led us to consider it as a likely runaway star. We summarize the basic data of HD~137071 in Table~\ref{basicpars}, where astrometric data are taken from the Gaia DR2 catalog, radial velocity from \citet{Famaey05}, and photometry from the 2MASS All-Sky Catalog \citep{Skrutskie06}. The distance is taken from the parallax-to-distance conversion of \citet{Bailer18}.

\begin{table}[t]
\caption{Basic data of HD~137071}
\begin{tabular}{ll}
\hline
\smallskip\\
$\alpha (2000)$ & 15:22:37.37 \\
$\delta (2000)$ & +39:34:53.3 \\
$l$ & $64^\circ 425$ \\
$b$ & $+56^\circ 363$ \\
$\mu_\alpha \cos \delta$ (mas~yr$^{-1}$) & $1.525 \pm 0.181$ \\
$\mu_\delta$ (mas~yr$^{-1}$) & $-14.194 \pm 0.224$ \\
radial velocity (km~s$^{-1}$) & $-13.38 \pm 0.20$ \\
parallax (mas) & $1.5675 \pm 0.1212$ \\
distance to the galactic plane (pc) & $524^{+43}_{-38}$ \\
$J$ & $2.823 \pm 0.258$ \\
$H$ & $2.042 \pm 0.184$ \\
$K_S$ & $1.758 \pm 0.278$ \\
\smallskip\\
\hline
\end{tabular}
\label{basicpars}
\end{table}

\section{New observations and spectral classification\label{new_obs}}

We have observed HD~137071 in the course of an ongoing program to obtain homogeneous spectral classifications of a sample of suspected red supergiants within 1~kpc for the Sun. We used CAFOS, the visual imager and low-resolution spectrograph at the Calar Alto 2.2m telescope, on the night of 22/23 February 2019, with a setup consisting of a grating providing a resolution $\lambda / \Delta \lambda = 1800$ over the interval 6000-9000~\AA . We used the same setup to observe a grid of MK spectroscopic standards from the list of \citet{Garcia89} that included the giants HD~173780 (K2III), HD~164058 (K5III), HD~194193, (K7III) and the supergiant HD~216946 (K5Ib) as the closest spectral types for comparison with HD~137071, all of them with near-solar metallicity determinations in the literature. The wavelength range includes several temperature-sensitive TiO bands, the CaII infrared triplet, and numerous other atomic and molecular features useful for the determination of spectral subtype and luminosity class.

\begin{figure}[ht]
\begin{center}
\hspace{-0.5cm}
\includegraphics [width=9cm, angle={0}]{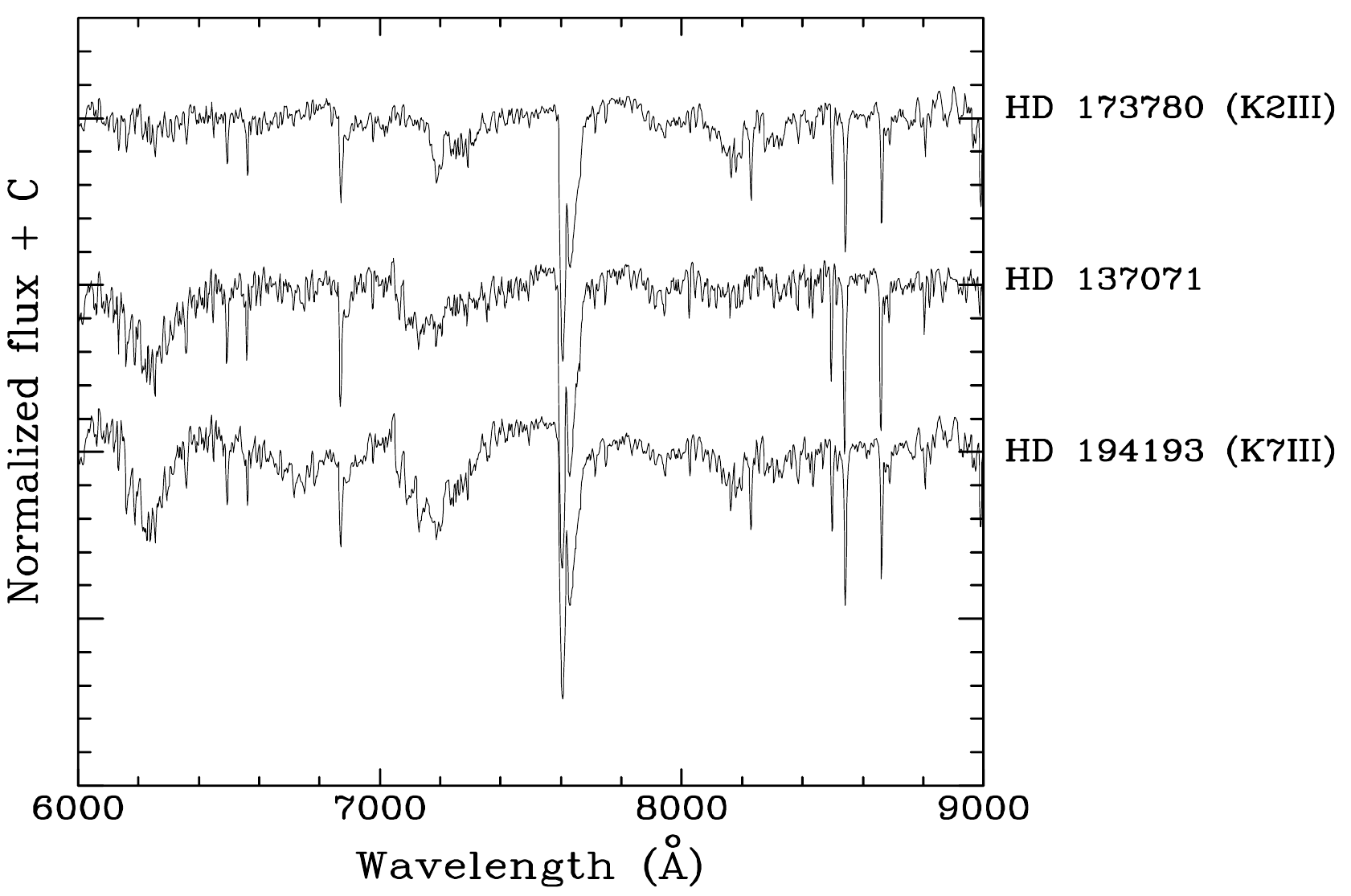}
\caption []{A comparison of the spectrum of HD~137071 with those of two luminosity class~III MK standards bracketing its spectral subtype.}
\label{comp_classIII}
\end{center}
\end{figure}

\begin{figure}[ht]
\begin{center}
\hspace{-0.5cm}
\includegraphics [width=9cm, angle={0}]{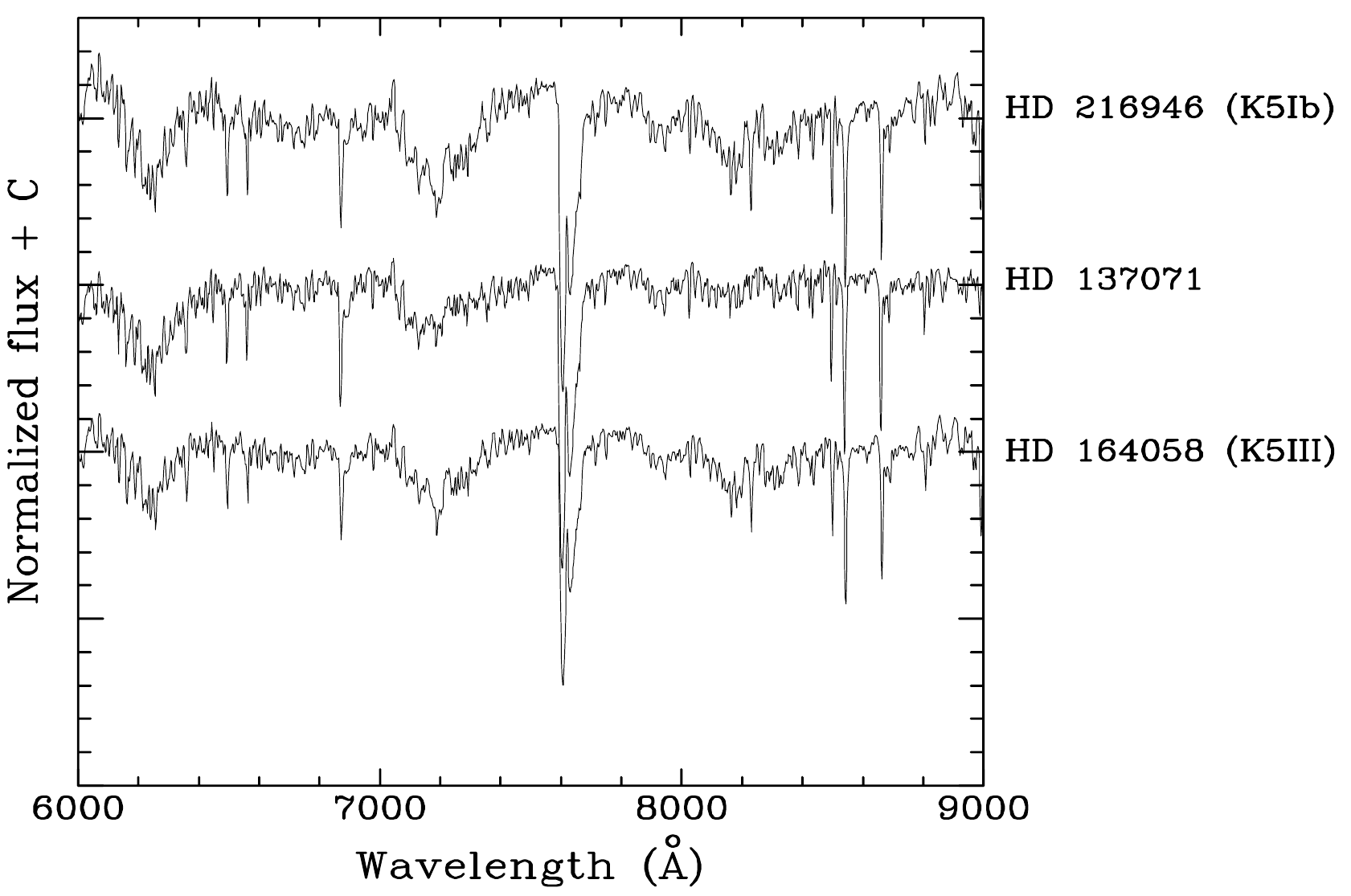}
\caption []{A comparison of the spectrum of HD~137071 with those of two K5-type MK standards of different luminosity class.}
\label{comp_K5}
\end{center}
\end{figure}

Figure~\ref{comp_classIII} compares the spectrum of HD~137071 with that of the K2III and K7III standards, clearly showing broad features that are intermediate between both, whereas Figure~\ref{comp_K5} compares the spectrum of HD~137071 with that of the two K5 standards of different luminosity classes. The overall appearance of the spectra, and particularly the depth of the broad temperature-sensitive TiO absorption feature centered around 7150~\AA\ ,  suggest a spectral type slightly earlier than K5, although definitely later than K2. We therefore classify HD~137071 as K4, in accordance with several previous classifications in the literature.

The assignment of a luminosity class is less straightforward, as it is based on subtler effects on luminosity-sensitive lines. A comprehensive set of features with positive luminosity effect (depth of the absorption feature increasing with luminosity), many of them  lying within our wavelength range, is provided by \citet{Torres93} based on spectra taken at a resolution $R \simeq 500$. The main one is the depth of the CaII triplet lines, which can be isolated from temperature effects once the spectral subtype has been determined from other temperature-sensitive indicators as we have done. The CaII triplet is known to be sensitive to metallicity as well \citep{Diaz89}, but we do not consider this to be a cause of concern here given the narrow range of metallicities spanned by HD~137071 and the MK standards used for comparison. \citet{Torres93} also note the usefulness as luminosity indicators of the blend of several atomic lines near 6500~\AA , the broad CN feature around 7970~\AA , and the blend of TiI lines 7345, 7358, and 7364 \AA . We have defined pseudocontinuum points and measured the equivalent widths of all these features in our spectra of HD~137071 and the standard stars, and have added a few other luminosity-sensitive lines, also with a positive effect, previously noted by \cite{Keenan45}: KI at 7699~\AA , FeI-TiI at 8468~\AA , and FeI at 8514~\AA\ and 8699~\AA . Because of the high signal-to-noise ratio of our spectra of such bright stars the equivalent widths are determined to a precision better than $\simeq 5$\% for lines with equivalent widths above $5$~\AA \ and $\simeq 10$\% for the weaker lines, the uncertainty being dominated by the exact location of the pseudocontinuum adopted for each line.

\begin{figure}[ht]
\begin{center}
\hspace{-0.5cm}
\includegraphics [width=9cm, angle={0}]{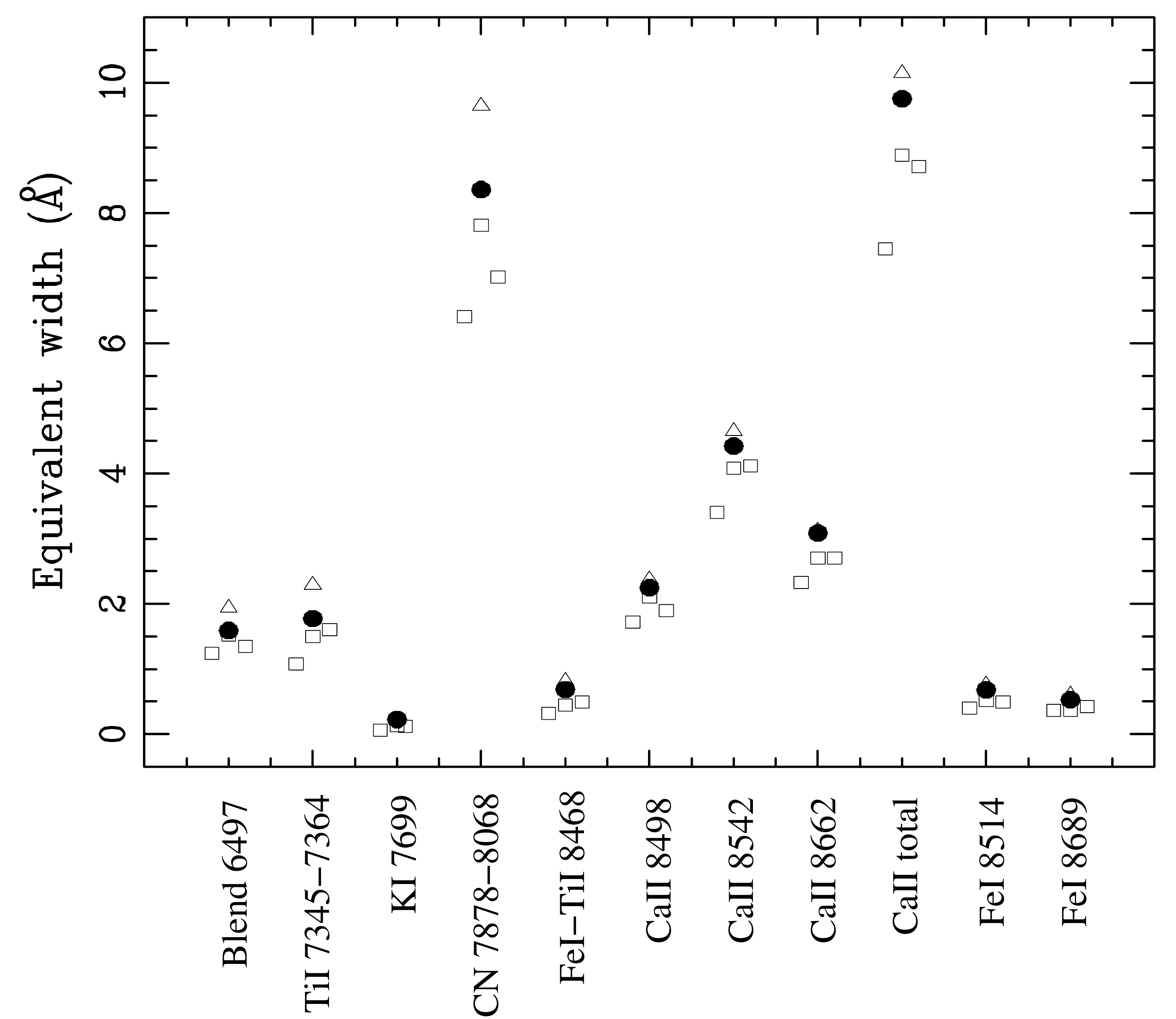}
\caption []{Equivalent widths of luminosity-sensitive features in the spectrum of HD~137071 (filled circles) compared with those of three luminosity class III giants (open squares) and the luminosity class Ib supergiant HD~216946 (open triangles). Three open squares are plotted for each feature, corresponding from left to right to HD~173780 (K2III), HD~164058 (K5III), and HD~194193 (K7III).
}
\label{indexes}
\end{center}
\end{figure}

The results are summarized in Figure~\ref{indexes}, where the equivalent widths of all these features measured on the spectrum of HD~137071 are compared with those of the two K5 standards. The equivalent widths of all the features without exception in the spectrum of HD~137071 fall between those of the K5Ib supergiant and the K5III giant. We also show our measurements of those same features on the spectra of the K2III and K7III standards, which also fall consistently below those of HD~137071 thus ruling out a slight spectral subtype misclassification as the possible cause of the systematic differences. Based on this, we propose K4II as the definitive spectral classification of HD~137071.

We also observed HD~137071 with the AstraLux lucky imager \citep{Hormuth08} at the Calar Alto 2.2m telescope on the night of 10~March~2020. We obtained four sequences of 10,000 images in the $V$ filter, each with individual exposure times of 30 milliseconds, stacking the 1\% best images of each sequence. The resulting images have a point-spread function (PSF) with a sharp core having a full width at half-maximum of $0''28$, which allows us to search for relatively close companions. We have estimated detection limits by adding to the actual image a number of simulated stars of different magnitudes at various position angles and distances, each with the PSF of the real star scaled accordingly to the chosen magnitude of the simulated star. We then perform photometry at the position of the simulated stars, deriving in this way detection limits at arbitrary angular distances from HD~137071. In this way, we can exclude the presence of a physical main sequence companion to HD~137071 with $V < 7.5$ (corresponding to a spectral type earlier than B2.5V) at more than $0''28$, corresponding to a projected physical distance of 180~pc. Fainter companions are also ruled out at progressive larger separations, down to $V=11$ (the approximate sensitivity limit of our AstraLux observations) at distances larger than $1''2$.

\section{Results and discussion\label{results}}

\subsection{Physical parameters\label{physpar}}

Using the temperature vs. spectral type calibration of \citet{Levesque07} for galactic supergiants we estimate the effective temperature of HD~137071 to be $T_{\rm eff} = 3900$~K, in excellent agreement with the value $T_{\rm eff} = 3892$~K obtained from the atmosphere model fit by \citet{Prugniel11}. We obtain the intrinsic $(J-K)_0$ color in the Johnson-Glass photometric system from \citet{Kucinskas05}, using their models for solar metallicity and $\log g = 1.0$. The latter choice is based on the value of $\log g$ of HD~137071 from \citet{Prugniel11}, although the precise value is of marginal importance at most given the mild dependence of the intrinsic colors in the relevant range of $\log g$. The $(J-K_S)_0$ color in the 2MASS system is obtained using the transformation $(J-K_S)_{\rm 2MASS} = 0.983 (J-K)_{\rm JG} -0.018$ \citep{Carpenter01}. The value obtained, $(J-K_S)_0 = 0.93$, is slightly lower than the actual $(J-K_S)$ color of HD~137071, suggesting the existence of moderate extinction in its direction. Using the extinction law of \citet{Cardelli89}, we obtain $A_K = 0.686 [(J-K_S) - (J-K_S)_0] = 0.093$ ($A_V = 0.8$), having a virtually negligible impact on the luminosity that we derive from the infrared photometry.

The distance estimated by \citep{Bailer18} from the Gaia parallax translates into a distance modulus $DM = 8.99 \pm 0.17$. Using the $K-$band bolometric correction vs. spectral type for supergiants of \citet{Levesque07} we adopt $BC_K = 2.61$, which yields a luminosity

$$\log L(L_\odot) = -0.4 (K_S - DM - A_K + BC_K - 4.74) = 3.78 \pm 0.13 \eqno(1)$$

This is in good agreement with the luminosity class assigned from the spectral features discussed above. We note that some analyses caution about the use of Gaia DR2 parallaxes for bright stars as some issues are known to exist with systematics \citep{Lindegren18}, and  \citet{Drimmel19} note effects that can reach deviations as large as $1-2$~mas among stars brighter than $G \simeq 5$. At a magnitude $G = 4.86$, HD~137071 is therefore close to the faint end of the range where the effect be can become prominent. The data presented by \citet{Drimmel19} suggest that the effect is probably color-dependent, causing some bright red stars to have underestimated parallaxes. If this effect were present for HD~137071 to the same level as in the most extreme cases shown by \citet{Drimmel19}, its actual distance modulus could be reduced to as little as $8.0$ mag and its luminosity to $\log L(L_\odot) \simeq 2.8$, placing it within the range of normal giants with an initial mass of a few solar masses. However, the consistency of the luminosity-dependent spectral features with a luminosity class II rather than III show that this is not the case. Furthermore, the renormalized unit weight error, which is a measure of the goodness of the astrometric solution \citet{Lindegren18}, has a value of $1.1$ for HD~137071, below the threshold of 1.4 for which the astrometric solution is considered to be problematic. In any case the existence of poorly characterized systematics at bright magnitudes highlight the importance of the spectroscopic confirmation of the high luminosity presented here without relying solely on the published parallax measurement.

\begin{figure}[ht]
\begin{center}
\hspace{-0.5cm}
\includegraphics [width=9cm, angle={0}]{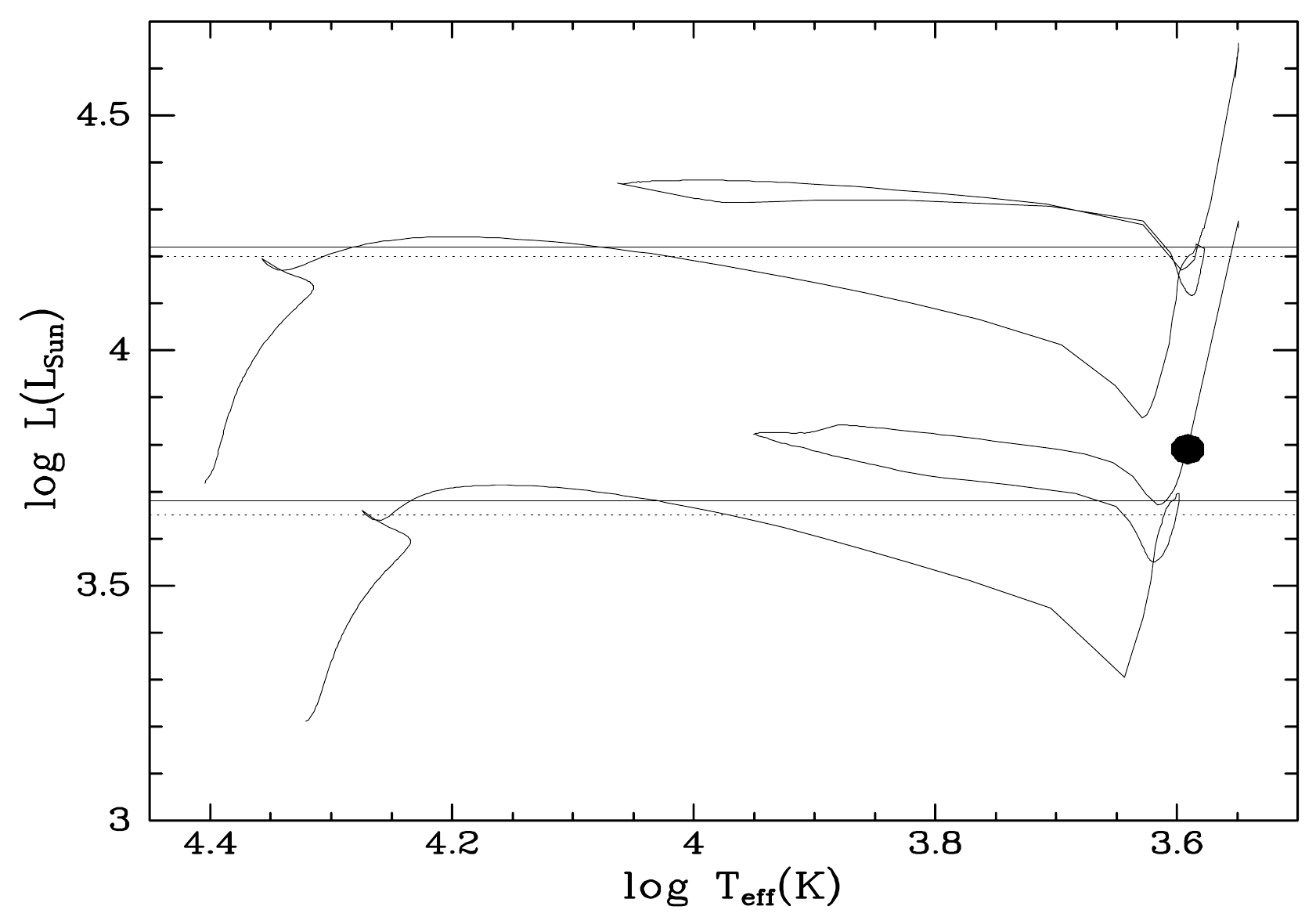}
\caption []{Evolutionary tracks for stars of 7~M$_\odot$ (lower curve) and 10~M$_\odot$ (upper curve) initially rotating at 60\% of the critical velocity, from \citet{Ekstroem12}. The horizontal solid lines indicate the luminosity of the star at the beginning of the He-burning phase for the rotating case illustrated in the figure, and the dotted lines show the same for the non-rotating models. Although the luminosity, and therefore the derived mass, is only weakly dependent on the initial rotation velocity, the age at which the He-burning stage is reached is significantly older for a rotating star as compared to the non-rotating case.}
\label{tracks}
\end{center}
\end{figure}

We assume that HD~137071 is undergoing the He-burning phase before entering the blue loop, as the post-blue loop phase ending with the supernova explosion is much shorter. A comparison of the values of $T_{\rm eff}$ and $\log L$ in the relative stable He-burning phase with the evolutionary tracks for massive solar metallicity stars \citep{Ekstroem12} indicates an initial mass of 8~M$_\odot$, with a small dependency on the initial rotation velocity of the precursor. Figure~\ref{tracks} shows the location of HD~137071 between the evolutionary tracks for stars of 7~$_\odot$ and 10~M$_\odot$, with the effect of rotation schematically indicated at the beginning of the He-burning phase.

The age has nevertheless a stronger dependency on the assumed rotation velocity of the precursor. At the initial mass of 8~M$_\odot$, the evolutionary models of \citet{Ekstroem12} show that the start of the He-burning phase varies considerably as a result of the mixing enhancement effect in the stellar interior caused by rotation. In this way, a supergiant with a non-rotating precursor enters the He-burning phase at the age of $\sim 35$~Myr, increasing to $\sim 42$~Myr if the precursor rotates at 60\% of the critical velocity. The duration of the pre-blue loop He-burning phase (effectively the time spent by the star in the red supergiant phase) is in both cases $\sim 3$~Myr. The available data do not allow us to adopt a more precise value of the age, but even such a wide range places gives some useful constraints on the origin of HD~137071 as a runaway star.

\subsection{Kinematics, birthplace, and the origin of HD~137071 \label{kinematics}}

The runaway nature of HD~137071 is revealed solely by its position and kinematics in combination with its youth. The current peculiar velocity with respect to the local standard of rest is $54.1$~km~s$^{-1}$, largely in the directions parallel to the galactic plane, with a very small vertical component ($2.1$~km~s$^{-1}$) at present. Unlike the case of the more massive galactic M-type supergiant runaways mentioned in the introduction, no arc-like structure tracing a bow shock is identified in the proximities of HD~137071 that would result from the interaction of its stellar wind with the interstellar medium, and that we would have expected to appear in infrared images of the WISE All-Sky Atlas if present. This may be due both to the weaker winds from red supergiants descending from B-type stars, as well as to the highly rarified interstellar gas at its distance from the galactic plane. Unlike most runaway stars known, the proper motion and radial velocity of HD~137071 indicate that it is moving toward the galactic plane, rather than away from it. Given its youth and the considerable distance to the galactic plane, this most likely indicates that HD~137071 is making its first return toward the galactic plane after having been born in its proximity and been subsequently ejected.

To explore the past trajectory of HD~137071 we have used the GRAVPOT16 web utility\footnote{https://gravpot.utinam.cnrs.fr/}, based on the Besan\c{c}on Galaxy model \citep{Robin12} as described in the PhD Thesis of J.G. Fern\'andez-Trincado (2017) and leaving as default parameters those in \citet{Helmi18}. The reconstruction of the galactic orbit of HD~137071 under this potential indicates that the star reached its largest distance from the galactic plane only 2~Myr ago, as indicated by the very small value of the component of the velocity perpendicular to the galactic plane. Tracing the trajectory further back in time, HD~137071 would have been on the galactic plane $31.5\pm 0.5$~Myr ago. At that point, the components of the velocity with respect to its local standard of rest were $U = 20.6$~km~s$^{-1}$ (away from the galactic center), $V = -32.9$~km~s$^{-1}$ (in the direction opposite to galactic rotation), and $W = 28.6$~km~s$^{-1}$ (toward the North galactic pole). Each of those quantities is twice or more higher than the velocity dispersion of the extreme Population~I in the corresponding coordinate, and imply a spatial velocity of $48.2$~km~s$^{-1}$ with respect to the local standard of rest at the time of the galactic plane crossing. Assuming that the birthplace of HD~137071 lies approximately at the location of its intersection with the galactic plane $31.5$~Myr ago, and that it has no peculiar motion with respect to the galactic circular rotation, its present location should be in the direction of galactic longitude $l \sim 93^\circ$, at a distance of $\sim 1.4$~kpc from the Sun. The open cluster NGC~7086 lies at that approximate distance \citep{Cantat18} about one degree away, but at an estimated age of $100$~Myr \citep{Rosvick06} it can be ruled out as the birthplace of HD~137071. No other candidate clusters or aggregates are known in the region. Our results remain essentially unchanged using the galactic potential of \citet{Mcmillan17}, which sets the galactic plane crossing at the slightly later time of 30~Myr ago.

On the other hand, any age within the possible range discussed in Section~\ref{physpar} implies that HD~137071 was already in existence at the time of the galactic plane crossing, then as a B2V main sequence star on its way to becoming a red supergiant about 30~Myr later. Taken at face value, even the lower limit to the age derived in Section~\ref{physpar} would situate the birthplace of HD~137071 on the opposite side of the galactic plane, at least 100~pc from it, and up to $\sim 400$~pc if the precursor were a longer-lived fast rotator. Those are rather large values when compared with the scale height of the thin disk traced by the extreme Population~I, and lead us to favor an interpretation in which the precursor of HD~137071 would have been instead born near the galactic plane, becoming a runaway only after having spent several Myr on the main sequence. Such a lag, perhaps reaching as many as $\sim 20$~Myr, between the formation of HD~137071 and its ejection as a runaway tends to favor the supernova disruption mechanism rather than the dynamical interaction one as the cause of its runaway nature. Ejection of most stars ending up as runaways is expected to take place in the initial relaxation phases of the dense, moderately massive clusters that simulations suggest as the preferred scenario for dynamical ejection \citep{Fujii11}. As a consequence, we would have expected the time since ejection from the cluster to be very similar to the age of the star, which does not seem to be the case for HD~137071.

If the runaway velocity of HD~137071 instead has its origin in the supernova explosion of a more massive companion, the ejection velocity derived above essentially reflects the orbital velocity of HD~137071 at the time of the supernova explosion of the companion, which allows us to estimate the possible range of initial separations between both. Taking $\sim 50$~Myr as an upper limit to the age of HD137071, corresponding to the extreme case of a precursor rotating at near-critical velocity, and assuming that the ejection took place $31.5$~Myr ago near the galactic plane, the difference between the age of HD~137071 and the time of the ejection event yields an upper limit to the lifetime of the companion, hence a lower limit to its mass, and a lower limit to their separation assuming a circular orbit at the time of the explosion. The companion lifetime of $\sim 18.5$~Myr implies an initial mass $\sim 11$~M$_\odot$, and the combined masses of HD~137071 and its companion, together with the orbital velocity $48.2$~km~s$^{-1}$, yield a semimajor axis of their orbit of $2.4$~AU. Conversely, assuming a massive companion of $22$~M$_\odot$, the highest mass for which a supernova explosion is the most likely outcome rather than direct collapse into a black hole \citep{Sukhbold16}, the semimajor axis would be $6.1$~AU. Those values are comparable or even below the radius of the massive companion in the pre-supernova red supergiant phase, implying that HD~137071 should have undergone a period of intense accretion through Roche lobe overflow of its companion, and perhaps a common envelope phase, besides the subsequent contamination by the supernova ejecta.

\section{Concluding remarks \label{conclusion}}

HD~137071 has the temperature, luminosity and spectral characteristics of a low-mass K-type supergiant, and the kinematics of a runaway star, being to our knowledge the only star of this type identified thus far. Statistically, the fraction of runaway stars drops from $\sim 25$\% to $\sim 5$\% from O to B types \citep{Stone91}, but the decreasing fraction should be offset by the rising initial mass function with decreasing masses. K-type supergiant runaway stars should therefore be abundant in absolute numbers, and the improved distances being obtained with Gaia should allow the identification of many more. Nevertheless, the caveat subsists that convective activity in their surfaces and their large radii ($\simeq 0.9$~AU for a supergiant with a $8$~M$_\odot$ precursor, like HD~137071, and even larger for more massive precursors) sets limits to the accuracy with which even Gaia can determine their parallaxes, making it very advisable to spectroscopically confirm the supergiant nature of candidate runaway stars to avoid samples contaminated with older, non-runaway red giant branch stars. Reliable samples of K supergiant runaway stars obtained in this manner should make it possible in the future to reconstruct the history of ejections by either of the proposed mechanisms in the solar neighbourhood back to a time hardly accessible through main sequence O and B stars.

\begin{acknowledgements}

It is always a pleasure to thank the hospitality of all the staff at Calar Alto Observatory, and particularly David Galad\'\i\ on this occasion for his support with our observations. This work was supported by the MINECO (Spanish Ministry of Economy) through grant RTI2018-095076-B-C21 (MINECO/FEDER, UE). The Two Micron All Sky Survey (2MASS) is a joint project of the University of Massachusetts and the Infrared Processing and Analysis Center/California Institute of Technology, funded by the National Aeronautics and Space Administration and the National Science Foundation. This research has made use of the SIMBAD database, operated at CDS, Strasbourg, France.

\end{acknowledgements}

\bibliographystyle{aa} 
\bibliography{HD137071_cit}





\end{document}